\documentclass[10pt,letterpaper]{article}
\usepackage{opex3}

\usepackage{graphicx}   
\usepackage{algorithm}
\usepackage{algpseudocode}
\usepackage{hyperref}   

\begin{document}

\title{Fast algorithms for generating binary holograms}

\author{Dustin Stuart, Oliver Barter and Axel Kuhn$^*$}
\address{Clarendon Laboratory, University of Oxford, Parks Road, Oxford, OX1 3PU, UK}
\email{$^*$axel.kuhn@physics.ox.ac.uk}

\begin{abstract}
We describe three algorithms for generating binary-valued holograms. Our methods are optimised for producing large arrays of tightly focussed optical tweezers for trapping particles. Binary-valued holograms allow us to use a digital mirror device (DMD) as the display element, which is much faster than other alternatives. We describe how our binary amplitude holograms can be used to correct for phase errors caused by optical aberrations. Furthermore, we compare the speed and accuracy of the algorithms for both periodic and arbitrary arrangements of traps, which allows one to choose the ideal scheme depending on the circumstances.
\end{abstract} 

\ocis{(350.4855) Optical tweezers or optical manipulation, (090.1760) Computer holography, (090.1000) Aberration compensation}


\section{Introduction}

Since their invention by Arthur Ashkin \cite{Ashkin:86}, optical tweezers have had an enormous impact in diverse fields from biology to quantum physics (see \cite{Ashkin2000} for a review). The underlying mechanism is the optical gradient force, which acts on polarisable particles such as living cells \cite{enlighten95082}, nanoscale tools \cite{enlighten62628} or single atoms \cite{PhysRevX.4.021034}, causing them to be trapped at the point of highest intensity of a tightly focussed light beam. The potential energy of such a particle is proportional to the intensity of the trapping light at the position of the particle. Much effort has been devoted to designing dynamic potential landscapes for trapping and moving large numbers of particles. Regular arrays of thousands of potential wells have been created using optical lattices \cite{NatPhys.3.556, Nature.462.74}, microlens arrays \cite{PhysRevLett.105.170502} and diffractive optical elements \cite{PhysRevA.88.013420}, however, these methods are limited in that the trapping sites can only be moved in unison, not individually. A second approach is to use an acousto-optic deflector (AOD) to generate a steerable trapping beam which can be used to `paint' time-averaged potentials \cite{NewJPhys.11.043030, 0953-4075-41-21-211001, Trypogeorgos:13}. In one demonstration \cite{Opt.Lett.39.2012} the authors generate 32 movable trapping beams using frequency shift key modulation. However, the time averaging is only valid when the oscillation frequency of the trapped particles is much lower than the rate of frequency shifting, which is ultimately limited by the rise time of the AOD.

\begin{figure}[tb]
\centering
\includegraphics{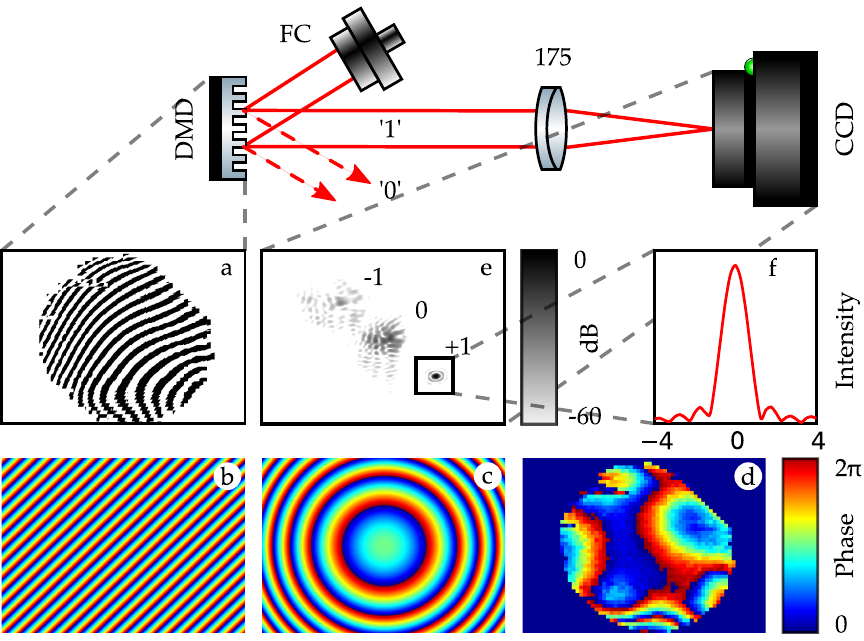}
\caption{The experimental setup used for testing the holographic optical tweezers. A collimated beam of laser light at 785 nm is delivered from a fibre coupler (FC) and is incident on the DMD. The reflected light from mirrors in the `1' state is focussed using an achromatic doublet lens (f = 175 mm). The traps are produced in the focal plane of this lens and imaged with a CCD camera. (a) The binary hologram on the DMD, formed from (b) a linear phase ramp, (c) a quadratic lens and (d) an arbitrary phase correction map. (e) The intensity measured by the CCD camera. (f) A line profile through the desired trap showing diffraction-limited performance (the $x$-axis is in dimensionless units of $\lambda D/f$ where $D$ is the diameter of the limiting aperture).}
\label{fig:setup}
\end{figure} 

A third, more flexible approach is to use a spatial light modulator (SLM) to create the desired potential landscape. The SLM can either be used as a variable attenuator to create the desired intensity pattern in the trapping light which can then be directly imaged on to the particles \cite{NewJPhys.14.073051}, or to display a hologram which is converted into the desired intensity landscape after propagation through a lens \cite{RevSciInst.83.11}. The direct imaging technique is simple but does not make efficient use of optical power as only a small area of the SLM is active for each trap. The holographic technique concentrates a large fraction of the optical power in the active trapping sites and allows for three-dimensional positioning, but requires additional effort to calculate the hologram. There are two broad categories of SLMs: phase modulators such as ferroelectric (FLC) modulators \cite{JModOpt.51.2235} and liquid crystal displays (LCD); and amplitude modulators such as digital mirror devices DMDs \cite{Mirhosseini:13}. Holographic optical tweezers typically use a phase modulating LCD, coupled with an iterative phase retrieval algorithm such as the Gerchberg-Saxton algorithm \cite{Optik.35.237}, mixed-region amplitude freedom (MRAF) \cite{Pasienski:08}, offset MRAF \cite{SciRep.2.721}, or conjugate gradient minimisation \cite{2014arXiv1408.0188H}. Here we demonstrate the use of a DMD in a holographic optical tweezers apparatus, along with algorithms that are much faster than standard iterative phase retrieval. DMDs are extremely fast compared to all other available SLMs, with full frame rates up to 20 kHz. The DMD we are using at present in our laboratory (Texas Instruments DMD Discovery 1100) is an array of 1024$\times$768 micro-mechanical mirrors. Each mirror can be switched between two different angles, which we refer to as `0' and `1'. A mirror in the `1' position reflects light through the remainder of the optical setup whereas a mirror in the `0' position deflects light towards a beam stop, thus acting as a binary amplitude modulator.

In this paper, we show how to use a DMD to holographically generate large arrays of individually movable trapping sites. The paper is divided into three sections. Firstly, we describe several different algorithms for rapidly calculating artificial holograms for an amplitude-modulating SLM, and we show how to apply these to modulators that only permit binary amplitude modulation such as a DMD. Secondly, we show how the DMD can be used as a wavefront sensor to accurately measure and correct for optical aberrations. The ability to do this is essential for producing tightly focussed diffraction-limited traps. Finally, we compare the different algorithms on the basis of speed of convergence, efficiency of use of laser power and accuracy of the resulting trapping potentials.

\section{Computation of Holograms}

\begin{figure}[t]
\centering
\includegraphics{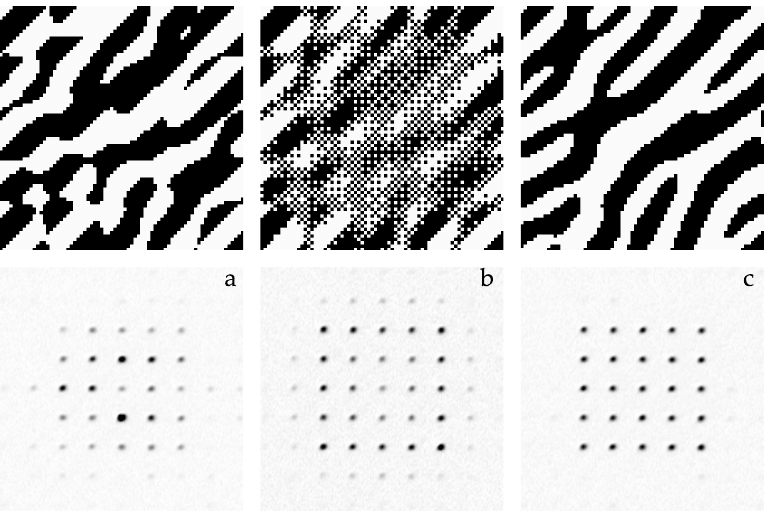}
\caption{Example images illustrating three of the  different hologram algorithms: (a) binary rounding, (b) ordered dithering, and (c) the Gerchberg-Saxton algorithm. Each case shows portion of the DMD ($73 \times 73$ px) showing the binary hologram and an image of the resulting intensity pattern taken with a CCD camera. The distance between the traps is 173 $\mu$m.}
\label{fig:sample}
\end{figure}

The general problem of holography is to produce an optical field $H(x,y)$, the hologram, which after propagation through a lens results in a desired optical field $F(x^\prime,y^\prime)$, the image. The problem can be divided in to two parts \cite{goodman1996introduction}: the computational problem of how to calculate the required optical fields; and the representational problem of how to display the complex-valued fields using a physical light modulator. We consider three different algorithms which from here on in we refer to as binary rounding, dithering, and weighted Gerchberg-Saxton. Figure \ref{fig:sample} shows examples of holograms and images generated using the different techniques.

\subsection{Hologram of a single focussed trap}

The common characteristics and limitations of all four algorithms are best illustrated by considering the hologram field of a single diffraction-limited trap at a precisely defined location:

\begin{equation}
H(x,y) = A \exp(i\frac{2\pi}{f \lambda}(x_0^\prime x + y_0^\prime y + \frac{z_0^\prime}{2f}(x^2 + y^2)) + i\phi(x,y))
\label{eq:hologram}
\end{equation}

\noindent $A = 1$ encodes the amplitude and phase of the trap, $x_0^\prime, y_0^\prime, z_0^\prime$ are its coordinates relative to the focal plane of the lens (with focal length $f$), and $\lambda$ is the wavelength of the trapping light. $x$ and $y$ are the coordinates in the plane of the DMD and $\phi(x,y)$ is an arbitrary function that we introduce to compensate for any systematic phase variations in the actual optical system. The terms in this equation are illustrated graphically in figure \ref{fig:setup}. The two terms linear in $x$ and $y$ (fig. \ref{fig:setup}b) allow the trap to be displaced within the focal plane of the lens. The $(x^2 + y^2)$ term (c) introduces a small defocus which allows the trap to be moved in and out of the focal plane by a distance $z_0^\prime$. Finally, the arbitrary phase correction (d) primarily compensates for the wave-front distortion due to the non-planar DMD surface we found in our demonstration setup. The image $F(x^\prime,y^\prime)$ may be calculated using the Fresnel diffraction formula \cite{goodman1996introduction}. If we neglect the quadratic lens term, this is essentially a Fourier transform of $H(x,y)$. The resulting $F(x^\prime,y^\prime)$ is a delta function spot at the location $(x_0,y_0)$. In reality, the spot is an Airy pattern whose width is determined by the limiting aperture of the setup (see fig. \ref{fig:setup}f). Furthermore, since only the intensity of the image $|F(x^\prime,y^\prime)|^2$ is relevant for trapping, we are free to choose the phase $\arg(A)$ of the trap.


\subsection{Binary rounding algorithm}

The simplest way to represent this complex-valued field on the DMD is the binary rounding algorithm: to map all pixels whose phase is between $-\pi/2$ and $\pi/2$ to the `1' state, and all other pixels to the `0' state. This simple mapping results in the maximum amount of optical power being directed in to the desired trap, since all the pixels in state `1' interfere constructively, and all the pixels that would interfere destructively are in state `0'. The result is a top hat grating (fig. \ref{fig:setup}a) where the fraction of the total optical power in the $n^{th}$ order of diffraction is given by $\frac12\mbox{sinc}(n\pi/2)^2$. The desired trap is produced in the $+1$ diffraction order, with a theoretical maximum power of $1/\pi^2 \approx 10.1\%$. The $-1$ and $0$ orders are blurred because the phase correction acts in the opposite direction for these orders - doubling the wavefront error.

\subsection{Extension to multiple traps}

It is straightforward to extend the above method to generating multiple traps by simply summing the complex-valued holograms for each trap. The $A$ parameter is used to set the magnitude and phase of each trap. We set the magnitude of the $n^{th}$ trap $|A_n| = 1/\sqrt N$ in order to evenly distribute the total power amongst the $N$ traps. In addition, we set the phase of each trap to a random value between $0$ and $2\pi$ in order to prevent the amplitude maxima of all of the individual trap holograms from accumulating at the same position.

Now we consider how to rapidly compute this sum. One approach is to explicitly evaluate equation \ref{eq:hologram} for each of the $N$ traps at all $M$ pixels and sum the results. This approach was taken by Bowman et. al. \cite{CompPhysComm.185.1} for a phase modulator, using a graphics processing unit (GPU) to parallelise the calculation. For $N = 25$ traps and a $M = 512\times512$ pixel grid, they achieve a frame rate of between 60 Hz and 500 Hz depending on the choice of hardware and accuracy of the output. A disadvantage is that the computation time scales as $N.M$ and so increases linearly with the number of traps.

The approach we have taken is to implement the sum using a fast Fourier transform (FFT). The advantages of this approach are numerous. Highly optimised FFT libraries (such as FFTW \cite{FFTW05}) exist that run on ordinary computer hardware. Furthermore, the algorithm scales as $M\log M$, which is independent of the number of traps. Thirdly, if the number of traps is sparse such that $N \ll M$ then only the rows that actually contain traps need to be transformed, yielding a speedup by a further factor of two. The FFT method is particularly efficient if the lens and phase correction terms are the same for all traps, as these can simply be multiplied with the end result of the FFT. If different lens terms are desired, the lenses can be incorporated efficiently via a convolution. Benchmarking on a 2GHz Intel Core 2 Duo CPU gives a frame rate of 300 Hz, which is comparable to the fastest speeds hitherto achieved but with the advantage of being independent of the number of traps and not requiring any specialised hardware.

\subsection{Ordered dithering algorithm}

Unfortunately, for the case of multiple traps, the problem of representing the complex hologram as an amplitude-only one is more difficult. The simple binary rounding algorithm is highly non-linear, which results in additional unwanted ghost traps and drastic variations in intensities of the traps. The non-linearity arises from two sources: the quantisation of continuous pixel values to two binary values, and the fact that any pixels that are outside the range $(0,1)$ are truncated. The ordered dithering algorithm seeks to minimise this non-linearity by mimicking continuous greyscales on the binary DMD. Dithering works by alternating pixels between `0' and `1' with a high spatial frequency so that the average pixel value is approximately equal to a continuous value.

The first step in the algorithm is to create a $512\times512$ array in which to store the target image $F$. We set each entry corresponding to a trap to $\exp(i\theta_n)/\sqrt N$ where $\theta_n$ is a random phase. Next, we perform an inverse Fourier transform on the array to find the hologram $H$. After applying the quadratic lens and phase correction, we take the real part of the complex-valued hologram field. This step similar to Gabor's \cite{1948Natur.161..777G} original approach (of recording the interference pattern between the field and a uniform reference wavefront) since the interference term is proportional to $\Re (H(x,y))$. In addition, we scale and add an offset of $1/2$ since the real part can be both positive and negative while the DMD only accepts positive values. Finally, we truncate to $(0,1)$ and dither the result by comparing it to a fixed threshold matrix, specifically, an $8\times8$ Bayer matrix \cite{bayer}. Alternatively, the ordered dithering step can be replaced with an error diffusion dithering algorithm such as Floyd-Steinberg (see \cite{floyd} for a description of the algorithm). A pseudocode version is given below:

\begin{algorithmic}[1]
\Function{Ordered Dither}{$x_n,y_n$}
\State $F(x_n,y_n) \gets \exp(i\theta_n)/\sqrt N$
\State $H \gets $\Call{IFFT}{$F$}
\State $H \gets H.\exp(i\frac{z_0\pi}{f^2 \lambda}(x^2 + y^2) + i\phi(x,y))$
\State $H \gets $\Call{Real}{$(H + 1)/2$}
\State $H \gets $\Call{Clip}{$H,0,1$}
\State $H \gets H > threshold\_matrix$
\State \textbf{return} $H$ 
\EndFunction
\end{algorithmic}

\subsection{Weighted Gerchberg-Saxton Algorithm}

Another alternative algorithm that we consider is based on an iterative phase-retrieval algorithm: the weighted Gerchberg-Saxton algorithm. This algorithm exploits the phase freedom of the individual trap holograms in equation \ref{eq:hologram}. The algorithm is initialised with a target image of traps with equal amplitudes and random phases. One then calculates the hologram by inverse Fourier transform and binary rounding. Next, the image that would be formed from this hologram is calculated by a forward Fourier transform. After this first iteration, the resulting traps in the image may have widely varied intensities. The initial image is updated according to the newly calculated image: the new phases are adopted and the intensity of the $n^{th}$ trap is weighted by $w_n$ such that weaker traps are boosted and more intense traps are attenuated. After a few dozen iterations, the algorithm converges towards a result with equal power in each trap. In pseudocode:

\begin{algorithmic}[1]
\Function{GerchbergSaxton}{$x_n,y_n$}
\State $F(x_n,y_n) \gets \exp(i\theta_n)/\sqrt N$
\Repeat
\State $H \gets $\Call{IFFT}{$F$}
\State $H \gets $\Call{Real}{$b$}$ > 0$
\State $F \gets $\Call{FFT}{$H$}
\State $w_n \gets w_n/abs(F(x_n,y_n))$
\State $F(x_n,y_n) \gets w_n. \exp(i$arg$(F(x_n,y_n)))$
\Until{convergence}
\State $H \gets $\Call{IFFT}{$F$}
\State $H \gets H.\exp(i\frac{z_0\pi}{f^2 \lambda}(x^2 + y^2) + i\phi(x,y))$
\State $H \gets $\Call{Real}{$b$}$ > 0$
\State \textbf{return} $H$ 
\EndFunction
\end{algorithmic}

\subsection{Wavefront correction}

Even though the DMD is an amplitude modulator, equation \ref{eq:hologram} allows for an arbitrary phase function $\phi(x,y)$ to be included in the hologram. This enables the correction of wavefront errors in the illuminating light. Such wavefront errors can be caused by optical aberrations in the lenses, or by the non-planarity of the DMD itself. The latter are particularly severe: we measured wavefront errors of up to $10\lambda$ over the surface of our particular DMD.

\begin{figure}[t]
\centering
\includegraphics{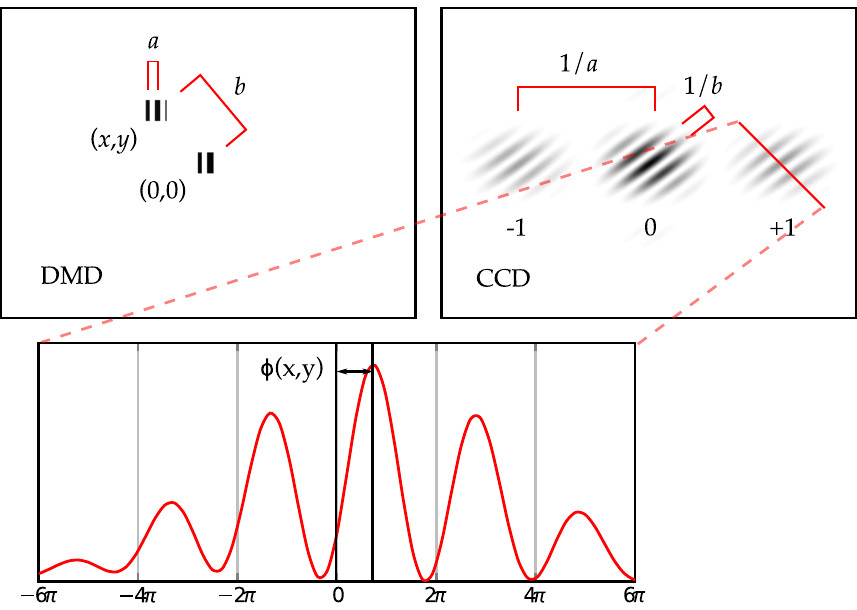}
\caption{Illustration of the technique for measuring wavefront errors (see text for description).}
\label{fig:correction}
\end{figure}

We determined the wavefront error at each point on the surface of the DMD using a technique similar to Cizmar et. al. \cite{Cizmar:2010p9920}. The technique is illustrated graphically in figure \ref{fig:correction}. The surface of the DMD is divided in to blocks of $16\times16$ pixels. All blocks were switched off except for two: one block of $16\times16$ pixels centred around the position $(x,y)$ and another centred at the origin. We denote the distance between the two blocks as $b$. The light from these two regions interferes on the CCD camera to form a sinusoidal Young's slit interference pattern whose fringe spacing is proportional to $1/b$. Within each block on the DMD, there is a diffraction grating with period $a$. This causes the $-1$, $0$ and $+1$ diffraction orders on the CCD to be separated by $1/a$. The fringes in the $+1$ order are examined by taking a line profile. The displacement of the interference fringe maximum from zero is proportional to $\phi(x,y)$, the phase of the light due to the wavefront error. We fit the interference pattern with a least-squares algorithm to extract this phase offset. This procedure is repeated for each block on the surface of the DMD, which takes approximately one minute. The resulting phase map is the one shown in figure \ref{fig:setup}d. Including this phase map in the hologram calculation eliminates the wavefront errors and yields a diffraction-limited spots in the first order of diffraction.

\section{Comparison of different algorithms}

\begin{figure*}[tb]
\centering
\includegraphics{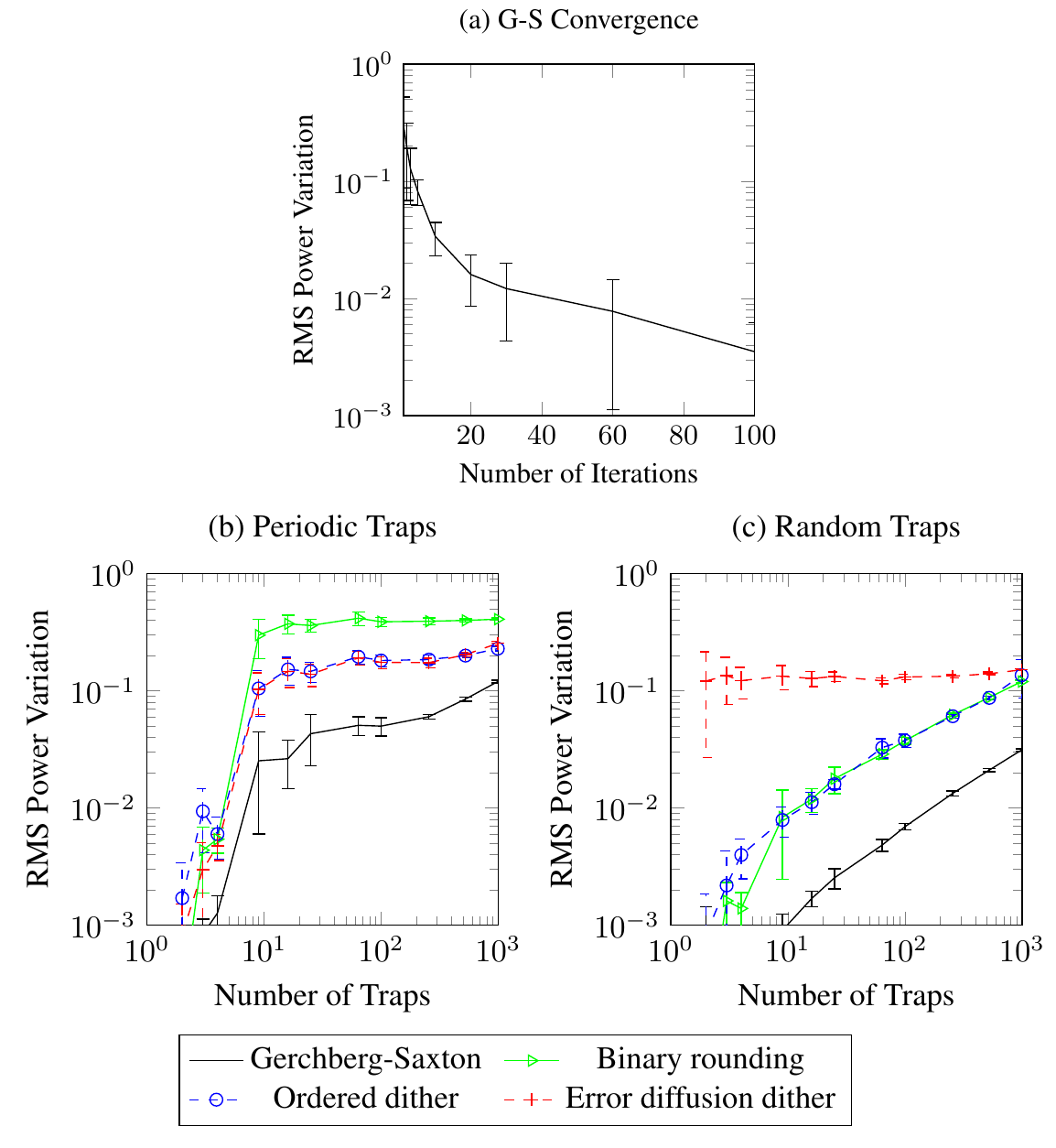}
\caption{Evaluation of the performance of the different algorithms for different numbers of traps. (a) The speed of convergence of the Gerchberg-Saxton algorithm for 25 periodic traps. (b) Relative RMS power variation for different periodic arrays of traps and (c) for randomly positioned traps. The error bars are statistical from a sample of 10 runs with different initial random phases.}
\label{fig:comparison}
\end{figure*}

All of the algorithms have similar capabilities. They can be used to position a large number of traps in three dimensions, and can all correct for wavefront errors. Furthermore, they all suffer from the same set of problems: loss of optical power in unwanted diffraction orders, generation of unwanted ghost traps, and variation of the power in existing traps. We evaluated the performance of each algorithm with a numerical simulation for two arrangements of traps: a periodic square lattice (as in fig. \ref{fig:sample}, spacing 4 pixels), and a set of random, uniformly distributed traps. The simulations were performed at a resolution of $512 \times 512$ pixels and averaged over 10 runs with different initial random phases for the traps. For the case of randomly positioned traps, their position was also randomised for each run. Figure \ref{fig:comparison} shows the results of how the severity of the power variation scales with the number of traps for each algorithm.

The loss of power originates from the conversion to a real-valued hologram. Both the binary rounding and Gerchberg-Saxton algorithm achieve the theoretical maximum of 10.1 \%, while the power per trap for the dithering algorithms is slightly lower. This is a consequence of line 5 of the ordered dithering algorithm, which can be expanded as

\begin{equation}
\frac12(\Re(H(x,y)) + 1) = \frac12 + \frac14H^*(x,y) + \frac14H(x,y)
\end{equation}

The $H(x,y)$ term is Fourier transformed to give $F(x^\prime,y^\prime)$, resulting in the desired arrangement of traps with a relative power of $1/4^2 = 1/16 = 6.25\%$, somewhat less than the theoretical maximum. Additionally, the $1/2$ term results in the zeroth-order diffraction spot containing $1/4$ of the total power, and the $H^*(x,y)$ term becomes $F(-x^\prime,-y^\prime)$, a mirrored version of the traps containing $1/16$ of the power.

The problem of ghost traps is not especially important for optical trapping as the ghost traps usually appear spatially separated from other traps, and are often too weak to trap particles. However, the problem of intensity variations is significant, and depends on both the arrangement of traps and the choice of algorithm. We quantify the variation by the relative RMS power variation i.e. the standard deviation of the power in each trap divided by the mean.

First, consider the general case of randomly positioned traps with random initial phases. The holograms of each trap add together incoherently. Therefore it is appropriate to add the intensities of each trap in quadrature, giving a total RMS amplitude of 1. After scaling according to line 5 of the ordered dithering algorithm, the majority of pixels in the hologram lie in the range $(0,1)$ and so the clipping step has a negligible effect on the traps. For random traps, even the binary rounding algorithm yields an acceptable results, with an RMS power variation of a few percent. The ordered dither algorithm achieves similar performance. Surprisingly, the error diffusion algorithm performs relatively poorly, especially for small numbers of traps. This is because the error diffusion dither distributes some noise to locations on the DMD which contain traps. In the case of ordered dithering, the noise is concentrated around a few specific locations corresponding to the spatial frequencies contained within the threshold matrix. Finally, the Gerchberg-Saxton algorithm is capable of reducing the RMS power variation even further. Figure \ref{fig:comparison}a shows how the RMS power variation improves with more iterations. High numbers of iterations suffer from diminishing returns. We choose 10 iterations as a compromise between speed and accuracy, but this is still 10 times slower than the ordered dithering algorithm.

However, for certain specific arrangements of traps, the binary rounding algorithm fails utterly. In the particular case of a periodic grids of traps, the maxima of the individual trap holograms tend to coincide at a few points on the DMD even with random initial phases. As a result, these locations in the hologram have an amplitude far outside the range $(0,1)$. The subsequent clipping step severely distorts the trap intensities. For periodic traps, the dithering algorithm is a significant improvement compared to the binary rounding algorithm. As before, the Gerchberg-Saxton algorithm can reduce the trap variation even further, but requires many iterations in order to do so.

For periodic arrangements of traps, the ordered dithering algorithm is a good compromise between power per trap, uniformity, and speed. When a periodic lattice of traps is not required, all algorithms provide good accuracy and the simple binary rounding algorithm is sufficient.

\section{Conclusion}

In conclusion, we have shown that a DMD can be used to holographically generate arrays of diffraction-limited light spots that are ideally suited for optical tweezers. While the efficiency is low (10.1\%), their fast full frame rate of 20 kHz is on the same timescale as the motion of laser-cooled atoms. Our tweezers have wide-ranging applications in the atomic physics wherever single atoms need to be manipulated accurately, for example, close to a dielectric surface, photonic crystal cavity, or superconducting circuit. In addition the high speed means that we can execute fast feedback on the atoms to precisely control their position. Our approach is not just limited to single atoms, but should enable new experiments in other fields (e.g. biological systems). Finally, the speed of each algorithm is crucial in order to utilise the full capability of the DMD. High speed algorithms such as ordered dithering are a significant simplification over iterative phase retrieval algorithms. High speed is necessary for real-time manipulation and feedback on trapped particles, and we have shown here that these can be realised using current technologies.

\section*{Acknowledgments}

The authors would like to acknowledge Lukas Brandt, Cecilia Muldoon, Edouard Brainis, Matt Himsworth and Jian Dong, whose contributions to experiments with non-holographic optical tweezers stimulated the development of this technique. DS would like to thank the Rhodes Trust for doctoral funding.

\end{document}